\documentclass[pra,twocolumn,longbibliography,amsfonts,amssymb,amsmath,floatfix,floats,a4paper]{revtex4-1}

\usepackage [margin=1in]{geometry}
\usepackage{physics}
\usepackage{graphicx}
\begin{document}

\title{Comment on ``Multifold paths of neutrons in the three-beam interferometer detected by a tiny energy kick"}

\author{Lev Vaidman}
\affiliation{Raymond and Beverly Sackler School of Physics and Astronomy, Tel-Aviv University, Tel-Aviv 69978, Israel}

\begin{abstract}
An  experiment with nested Mach-Zehnder interferometer [Phys. Rev. Lett. 111, 240402 (2013)] has been  recently  implemented with neutrons [Phys. Rev. A  97, 052111 (2018)]. Which-path information has been extracted from  faint traces the neutrons left, providing operational meaning to ``the particle's path''. The authors of the neutron interference experiment criticised the conclusions obtained by the authors of the optical experiment. I refute the criticism and argue that the results of the neutron interference experiment actually support the surprising picture of the past of the particle in the nested Mach-Zehnder interferometer.
\end{abstract}
\maketitle

The proposal to define where was a quantum particle in an interference experiment according to the faint traces it left \cite{past} generated a large controversy \cite{Danan,LiCom,RepLiCom,morepast,Jordan,Sali,SaliCom,Bart,BartCom,Poto,PotoCom,China,ChinaCom,Sok,SokCom,SokComRep,Grif,GrifRep,Nik,NikRep,NikRR,Hash,HashCom,HashComRep,Dupr,DuprCom,DuprComRep,Eli,Zhou,Disapp,Berge,BergeRep}. Most of the analyses made until now were about photons, but recently, multifold paths of neutrons in the three-beam interferometer were detected by observing a tiny energy kick \cite{Hasegava}. The neutron interference experiment was a slightly contracted version of an optical experiment \cite{Danan}. Although the experimental results of this neutron interference experiment and  the optical experiment showed similar structure, the authors of \cite{Hasegava} argued against the picture advocated in \cite{past,Danan}.

In the optical experiment two particular tunings of the nested Mach-Zehnder interferometer were considered, see Fig.~1. In the first run, there was a constructive interference   in both inner and  external interferometers, and faint traces were observed in all segments of the interferometer, $A$, $B$, $C$, $E$, and $F$, see Fig.~1a. In the second run, the inner interferometer was tuned to destructive interference toward the final beam splitter of the external interferometer and the faint traces were observed on path $C$, but also inside the inner interferometer on paths $A$ and $B$. No trace was observed on the way toward and out of the inner interferometer, segments $E$ and $F$, see Fig.~1b.

In the neutron interference experiment similar runs were performed, see Fig. 2. A small difference is that in the neutron interference experiment there was no segment $E$. To make an exact copy of the optical experiment  would require a crystal with five plates, instead of four which were used. (Even four plates is a technological achievement, since most previous neutron interference experiments had just three plates.)
\begin{figure}
\begin{center}
  \includegraphics[width=8cm]{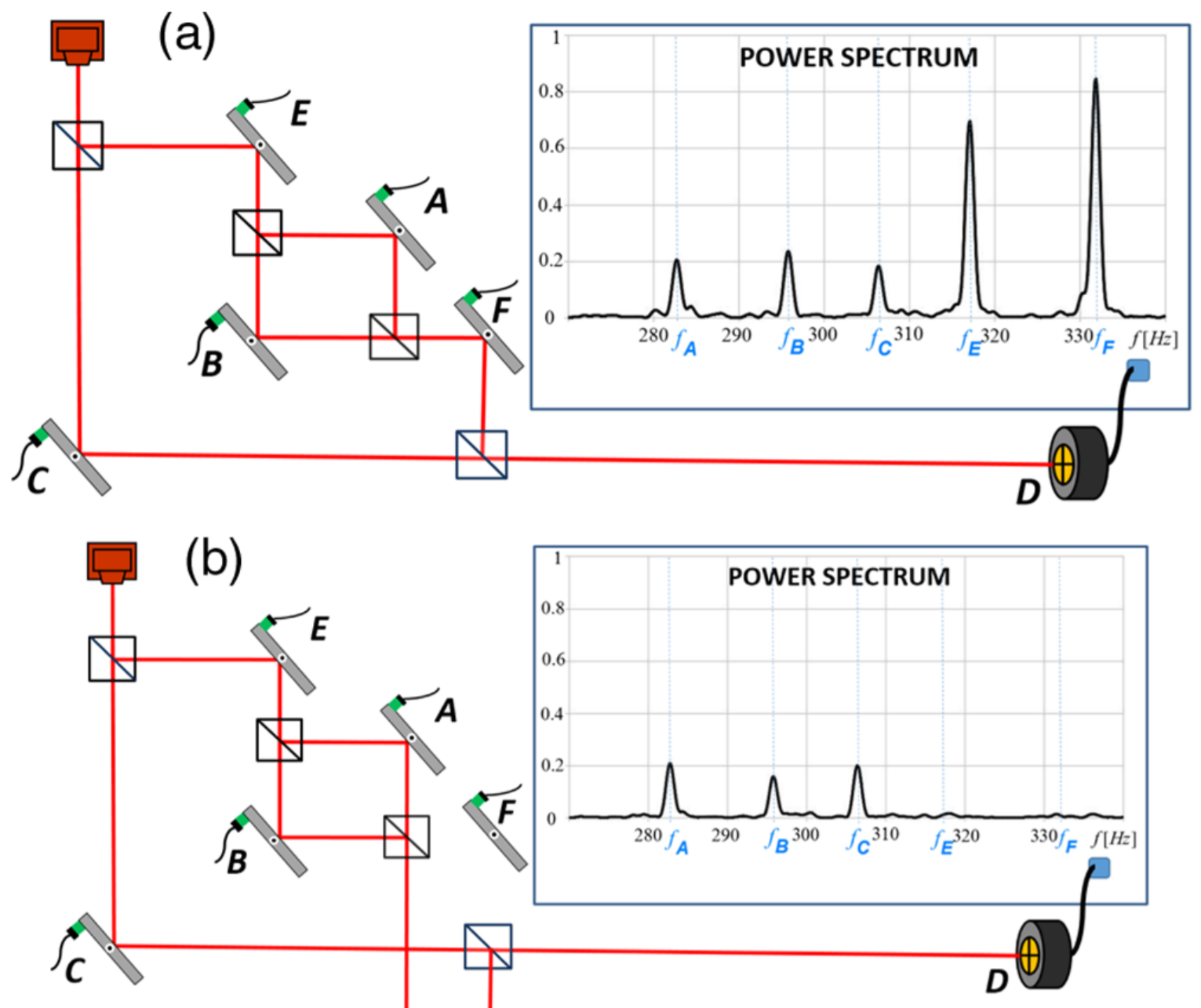}
\end{center}
  \caption{Measuring faint traces in optical nested Mach-Zehnder interferometer. a). Constructive interference. b) Destructive interference in the inner interferometer towards $F$. (Figures 3a and 3b in  \cite{Danan}).
  }
  \label{fig:FIG-1}
\end{figure}
\begin{figure}
\begin{center}
  \includegraphics[width=8cm]{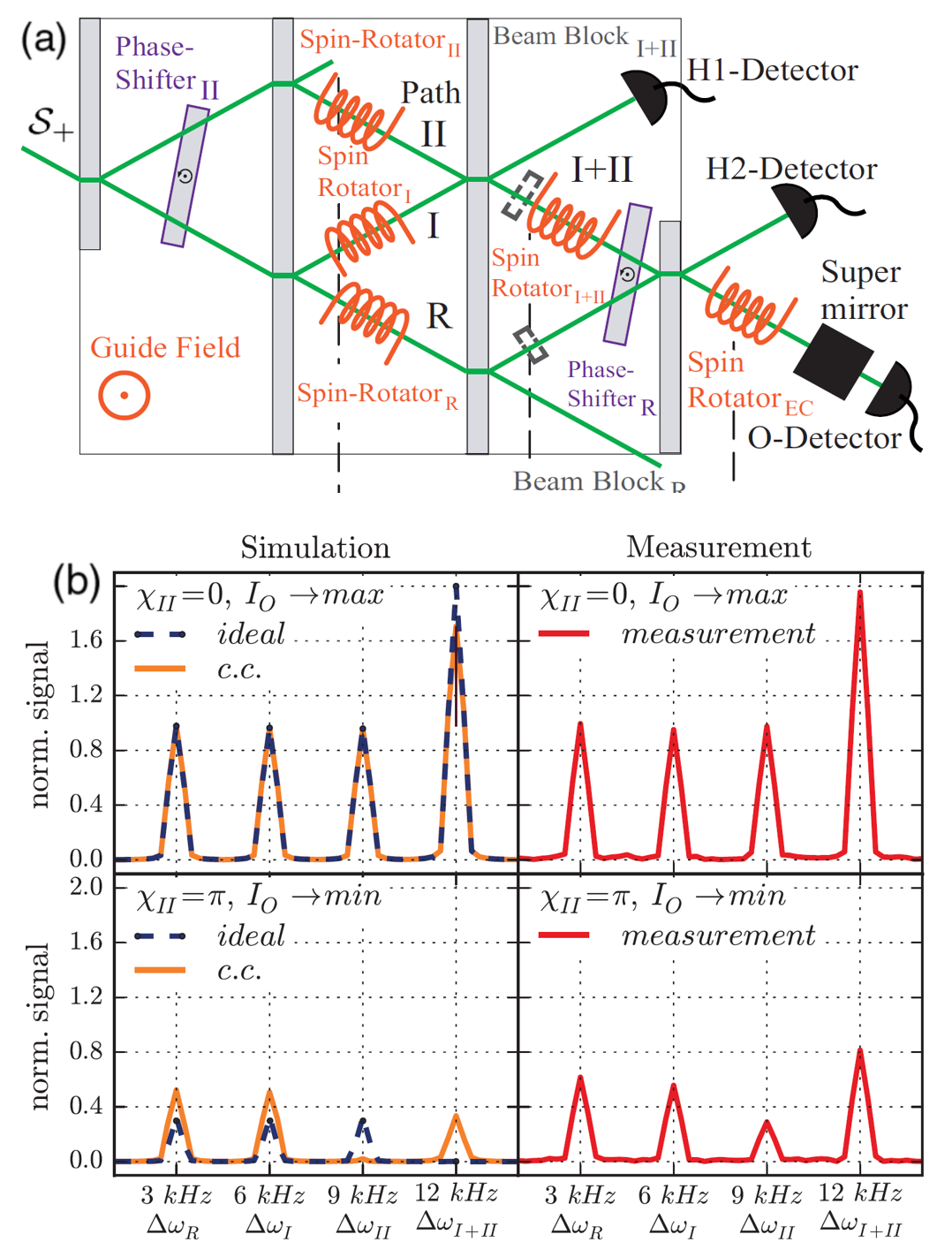}
\end{center}
  \caption{a). A schematic depiction of the experimental setup of the which-way measurement of neutrons in three-beam interferometer (Fig. 1a \cite{Hasegava}). The correspondence between paths of optical and neutron interference experiments: $A \leftrightarrow {\rm II}$, $B \leftrightarrow {\rm I}$, $C \leftrightarrow {\rm R}$, $F\leftrightarrow {\rm I+II}$. b) Simulation and actual results of the neutron interference experiments (Fig. 3a  \cite{Hasegava}). The first row was obtained for tuning  to constructive interference corresponding to optical experiment results in Fig. 1a. The second row was obtained for tuning  to destructive interference corresponding to  Fig. 2a. The relevant graphs, simulations for perfect contrast (the condition for the optical experiment)  are shown with dashed blue line.
  }
  \label{fig:FIG-2}
\end{figure}

In the run with complete constructive interference the results of the neutron interference experiment were similar to the results of the optical experiment: the traces were observed in $A$, $B$, $C$, and $F$. (There were no segment $E$). When the inner interferometer was tuned to destructive interference, the expected results were not obtained due to practical difficulty of obtaining perfect contrast, but the simulation of the expected results, as presented in blue dashed line of Fig. 2b showed traces in $A$, $B$ and $C$, but not in $F$. So, the experimental graphs (or at least theoretically expected graphs) were very much the same, although the particles and experimental methods were very different.

 What also was different is the interpretation. According to the ``faint trace'' criterion \cite{past}, places from which signals were obtained considered as places where the particle was, and places from which the signal did not arrive, in spite of existing coupling devices, corresponded to places where the particle was not present. The presence of the particle was {\it defined} by the trace it left. The authors of the neutron interferometer experiment disagreed:
 \begin{quote}
 ``... studying the interference effect, particularly in a (completely) destructive case, zero intensity appears; this situation is interpreted
in a mistaken manner as noncontinuous trajectories in Ref. [18]  (\cite{Danan} here). Appropriate consideration should be derived as the limit of (practically feasible) circumstances... There propagate (smaller and smaller numbers but still some) quantum particles; they are actually there.''
 \end{quote}

It is true that in any realistic interferometer and especially when we are trying to obtain (even weakly) the which path information, there will be no exact destructive interference. The question is: ``Where was a single neutron passing through the interferometer and detected by detector $O$?'' The fact that a small number of particles in  our pre- and postselected  ensemble actually were there, does not tell us that a particular neutron just detected was there.

Disregarding the observed faint traces  contradicts the whole point of the paper, the demonstration  of the past of the particle through the faint traces it left. It was stated in numerous places:

Title:
\begin{quote}
 ``... paths of neutrons in the three-beam interferometer detected by a tiny energy kick''.
   \end{quote}
   Abstract:
    \begin{quote}
    ``By ascertaining an operational meaning to ``the particle's path,'' which-path information is extracted from these faint traces with minimal perturbations.''
    \end{quote}
     End of Section II:
     \begin{quote}
     ``... which-way information is extracted from these oscillating intensities.''
 \end{quote}

More specifically, the passage related to the presence in $F$ (the path I+II) is:
 \begin{quote}
``WW information can be derived by a Fourier analysis of the time spectrum obtained at the $O$ detector. If
a Fourier component corresponding to a frequency $\Delta \omega_i$ is
found, this is clear evidence of neutrons having interacted with
the respective ${\rm SR}_i$. For instance, Eqs. (6) and (7) suggest that
neutrons have taken the paths I, II, and R for both settings but
that the path I+II has not been taken for the latter.''
 \end{quote}
  Both optical experiment \cite{Danan} and predictions of ideal neutron experiment tell us that the particle was not present in $F$. Another optical experiment, performed with single photons \cite{Zhou} also showed no signal from $F$, (see their Fig. 1b).

All results are consistent with the standard quantum theory. The question is the consistency of the definition of the past of a pre- and postselected quantum particle according the weak trace it leaves. The neutron interference experiment did not show anything mistaken about it.

The part of the optical experiment which was not reproduced with neutrons is testing the presence of the particle before it enters the inner interferometer, the arm $E$ in Fig. 1. It is somewhat unfortunate, because it could have been done with almost the same hardware, placing another spin rotator responsible  for measuring a faint trace in the lower beam, between the first and the second plate, see Fig. 2a. Obtaining surprising result that the particle was not in $E$,  requires preliminary tuning to destructive interference at detector $O$ with path II blocked.

Although the experiment has not been performed, the expected results were discussed in the paper:
\begin{quote}
``Rather different circumstances emerge when a WW marking is done on the beam, particularly before it enters the interferometer circuit.
... by tuning the phase setting at
the destructive interference position, i.e., the phase difference
of $\pi$, the WW signal does not emerge in the interfering
beam, say in the forward direction; the WW signal before the
interferometer is only redirected through the interferometer
loop. The situation that no WW signal is found in the final
detector does not necessarily mean that no WW signal has
existed at the position of the WW marking; these signals may
be redirected to other beams. This situation exactly happened in
an experiment by Danan et al. (see Fig. 3 in Ref. [18]  (Fig. 1b here)); the WW
signal at the mirror E is only redirected by the interferometer
circuit and does not arrive at the detector.''
\end{quote}

 In the approach \cite{past} the question was: ``Where was a pre- and postselected particle?'' Only the traces left by the particles detected by detector $D$ (or by the neutron detector $O$) should be considered. Since the measuring devices both in optical experiment \cite{Danan,Zhou} and in neutron interference experiment were the particles themselves, the signal could not be present in other beams. In the other beams were the particles with different postselection, which had  a different past and the records of a different past. A conceptually better experiment would use an external measuring device \cite{Steinberg}, but in any experiment we should consider the records conditioned on postselection in $D$ (or $O$). In all these cases there will be no signal from mirror $E$.

In summary,  the neutron interference experiment \cite{Hasegava} successfully observed the faint traces left by neutrons in some setups of the  three-path interferometer exhibiting results agreeing with similar  optical experiments \cite{Danan,Zhou}. However, the analysis in \cite{Hasegava} arguing  against the conclusions of \cite{past,Danan} was shown to be unfounded.

This work has been supported in part by the Israel Science Foundation Grant No. 1311/14 and
the German-Israeli Foundation for Scientific Research and Development Grant No. I-1275-303.14.



\end{document}